\documentclass{pasj00}
\draft

\begin{document}
\SetRunningHead{M.Sakai et al.}{Suzaku and XMM-Newton Observations of HESS J1614$-$518}

\title{Nature of the Unidentified TeV Source HESS J1614$-$518, Revealed by Suzaku and XMM-Newton Observations}

\author{
  Michito \textsc{SAKAI},\altaffilmark{1}
  Yukie \textsc{YAJIMA},\altaffilmark{1}
  and
  Hironori \textsc{MATSUMOTO}\altaffilmark{2}}
\altaffiltext{1}{Division of Particle and Astrophysical Science, Graduate School of Science, Nagoya University, \\Furo-cho, Chikusa-ku, Nagoya 464-8602}
\email{m\_sakai@u.phys.nagoya-u.ac.jp}
\altaffiltext{2}{Kobayashi-Maskawa Institute for the Origin of Particles and the Universe, Nagoya University, \\Furo-cho, Chikusa-ku, Nagoya 464-8602}

\KeyWords{acceleration of particles --- X-rays: individual (HESS J1614$-$518)} 

\maketitle

\begin{abstract}

We report on new Suzaku and XMM-Newton results concerning HESS
J1614$-$518, which is one of the brightest extended TeV
$\gamma$-ray sources and has two regions with intense
$\gamma$-ray emission. We newly observed the south and
center regions of HESS J1614$-$518 with Suzaku, since the
north region, including the position of the 1st brightest
peak of the TeV $\gamma$-ray emission, has already been
observed. No X-ray counterpart was found at the position of
the 2nd brightest peak of the TeV $\gamma$-ray emission; we
estimated the upper limit of the X-ray flux to be $1.6
\times 10^{-13}$ {\rm erg cm$^{-2}$ s$^{-1}$} in the 2--10
keV band. The soft X-ray source Suzaku J1614$-$5152,
which was found at the edge of the field of view in
a previous observation, was also detected at the middle of
HESS J1614$-$518. Analyzing the XMM-Newton archival data, we
revealed that Suzaku J1614$-$5152 consists of multiple point
sources. The X-ray spectrum of the brightest point
source, XMMU J161406.0$-$515225, can be
described by a power-law model with a photon index of $\Gamma
= 5.2^{+0.6}_{-0.5}$, or a blackbody model with
temperature $kT = 0.38^{+0.04}_{-0.04}$ keV. In
the blackbody model, the hydrogen-equivalent column
density is almost the same as that of the hard extended
X-ray emission, Suzaku J1614$-$5141, which
was found at the 1st peak position. If true, XMMU J161406.0$-$515225
may be physically related to Suzaku J1614$-$5141 and HESS J1614$-$518.

\end{abstract}

\section{Introduction}

Cosmic rays are the most energetic particles in the
Universe.  Since its discovery~\citep{Hes12}, their origin
has been a mystery.  A large fraction of the Galactic cosmic rays
reaching the earth are protons; only about 1\% of those are
electrons.

It has only been recently revealed that supernova remnants
(SNRs) may be effective electron accelerators, as the
detections of synchrotron X-rays originating from electrons
with energies approaching $E \sim 10^{14}$ eV (e.g.,
\cite{Koy95}).  However, a clue for proton accelerators ---
constituting the majority of cosmic rays --- evaded our
eyes.

Then the progress has been made.  The H.E.S.S. Cherenkov
Telescope, a high-sensitivity TeV $\gamma$-ray
telescope, has identified many very high energy (VHE)
$\gamma$-ray objects along the Galactic
plane~\citep{Aha05a,Aha05b}.  Most of the new objects are
spatially extended and therefore be Galactic objects, such
as pulsar wind nebulae (PWNe)~\citep{Ren08},
shell-type SNRs~\citep{Aha08}, and binaries~\citep{Aha05b}, etc.  
Nevertheless, the nature of
a large population of the VHE objects remains unknown.
These unidentified sources have no clear counterpart at
other wavelengths so that they are referred to as "dark
particle accelerators"~\citep{Aha05a}.  There are
two possible mechanisms to radiate the VHE $\gamma$-rays
efficiently: (1) electronic origin and (2) hadronic origin of
$\gamma$-rays.  In the former, VHE $\gamma$-rays results
from inverse-Compton upscattering of seed photons (cosmic
microwave background or other low-energy photons) by highly
energized electrons --- the same electrons which radiate
X-rays in the synchrotron process.  In the latter, VHE
$\gamma$-rays are explained by the decay of neutral pions
that originate in collisions between high-energy protons and
dense interstellar matter.  While TeV $\gamma$-ray emission
may allude to the presence of high-energy particles, it does
not differentiate the two possibilities.
However, the presence of X-ray emission (or a lack of
thereof) can shed a new light to the nature of these
mysterious $\gamma$-ray sources.

HESS J1614$-$518 (hereafter HESS J1614) is one of the
brightest, extended TeV $\gamma$-ray
sources~\citep{Aha05a,Aha06}.  It has two regions with
intense $\gamma$-ray emission.  Since it has no
viable counterpart identified in other wavelengths, HESS J1614 is
considered as a dark particle accelerator.  

\citet{Mat08} observed HESS J1614 in 2006 with Suzaku, an
X-ray astronomical satellite~\citep{Mit07}, to identify the TeV
$\gamma$-ray source in X-rays.  Two X-ray sources were found
in the 3--10 keV band.
One of the X-ray objects, Suzaku J1614$-$5141 (src A), is
extended and is located very close to the position of the
1st brightest peak of the TeV $\gamma$-ray emission.  The
other source, Suzaku J1614$-$5152 (src B), is located at the
middle of two TeV $\gamma$-ray peaks in HESS J1614.
Detailed analysis of src B was proven difficult with the
Suzaku observation in 2006 as the source resides near the
edge of the field of view and its flux estimate suffers
significant uncertainties.
Hence the second Suzaku observation was conducted in 2008 in
order to probe the nature of src B.
We also observed the position of the 2nd brightest peak of
the TeV $\gamma$-ray emission.

In the following sections, the results of the second Suzaku
observation will be discussed; in addition, our own analysis
on the XMM-Newton observation of the HESS J1614 region will
be utilized to do a spatial analysis and to examine temporal
behavior.  The uncertainties in described parameters are at
the 90\% confidence level; the errors in data points and
photon counts are at the 1$\sigma$ level, unless otherwise
stated.

\section{Observations and Data Reduction}

\begin{figure*}
  \begin{center}
    \FigureFile(120mm,80mm){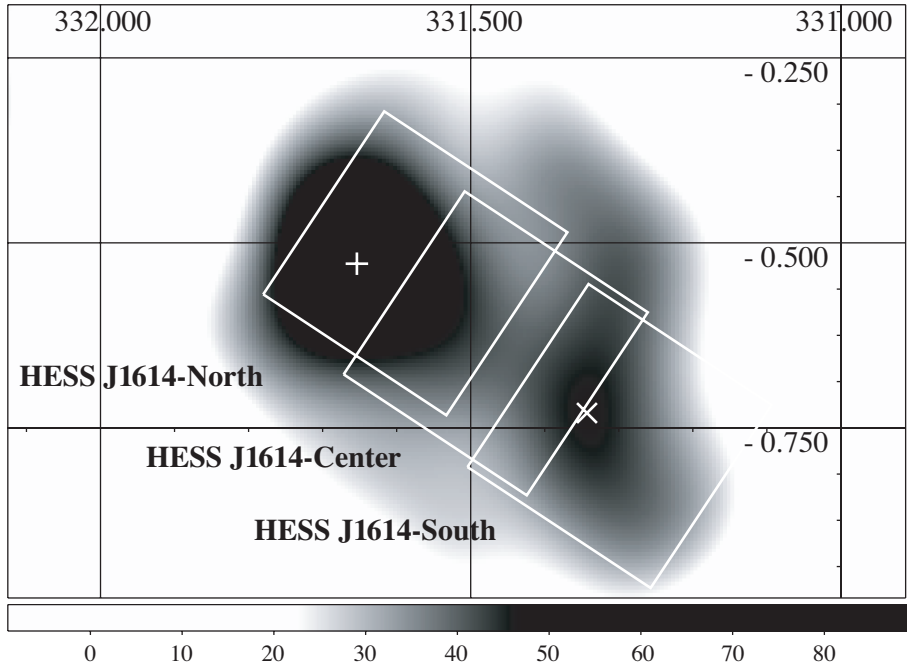}
  \end{center}
  \caption{Suzaku/XIS field of views (thick boxes) overlaid on the H.E.S.S. smoothed excess map. The scale bar below the figure represents the excess. The coordinates on the interior frame are Galactic. The plus and cross marks represent the positions of the 1st and 2nd brightest peaks of the TeV $\gamma$-ray emission.}\label{fig:1st}
\end{figure*}

The south and center regions of HESS J1614 were observed on
2008 September 20 and 21, respectively.  Figure
\ref{fig:1st} shows the Suzaku/X-ray Imaging Spectrometer
(XIS:~\cite{Koy07}) field of views in these
observations, together with that in the HESS J1614-North
observation performed on 2006 September 16.  The
observations are summarized in table \ref{tab:1st}.

Suzaku consists of two distinct co-aligned scientific
instruments.  One is the X-ray Imaging Spectrometer (XIS:~\cite{Koy07}),
which is an X-ray CCD camera located in
the focal plane of the X-ray Telescope (XRT:~\cite{Ser07}). 
The other is the Hard X-ray Detector (HXD:~\cite{Kok07,Tak07}),
which is a
non-imaging detector.  The observations were performed with
the three CCD cameras (XISs).  One of the XIS sensors (XIS
1) has a back-illuminated (BI) CCD, while the other two
(XISs 0 and 3) utilize front-illuminated (FI) CCDs.  One of
the FIs (XIS 2) suffered catastrophic damage on 2006
November 9, so that no useful data were obtained.  The XIS
was operated in the normal clocking mode (without the Burst
or Window options) with the Spaced-row Charge
Injection (SCI)~\citep{Uch09}.  The HXD data were also available.

We analyzed the data with the processing version of
2.2.11.22,\footnote{See
  $\langle$http:\slash\slash{}www.astro.isas.jaxa.jp\slash{}suzaku\slash{}process\slash{}history\slash{}v221122.html$\rangle$.}
utilizing the HEADAS software (version 6.8) and the
calibration database (CALDB) released on 2010 February 18.
All data affected by the South Atlantic Anomaly and
telemetry saturation were excluded.  We excluded the data
obtained with an elevation angle from the Earth rim of $<
5^{\circ}$.  Additionally, for the XIS data, we also
excluded the data obtained with that from the bright Earth
rim of $< 20^{\circ}$ and removed hot/flickering pixels.
After these data screenings, the effective exposures for the
XIS data were 53.7 ks and 47.7 ks, and those for the HXD-PIN
data were 40.9 ks and 43.0 ks, on the HESS J1614-South and
the HESS J1614-Center, respectively.

\begin{table*}[htbp]
  \begin{center}
    \caption{Log of Suzaku observations.}\label{tab:1st}
    \begin{tabular}{lcccccc}\hline
Name & OBSID & \multicolumn{2}{c}{Pointing direction} & Observation start & \multicolumn{2}{c}{Effective exposure (ks)}\\ 
  & & $l$ & $b$ & (UT) & XIS & HXD-PIN \\ \hline
HESS J1614-South & 503073010 & \timeform{331.2990D} & \timeform{$-$0.7611D} & 2008/09/20 18:18 & 53.7 & 40.9 \\
HESS J1614-Center & 503074010 & \timeform{331.4663D} & \timeform{$-$0.6358D} & 2008/09/21 13:31 & 47.7 & 43.0 \\ \hline
HESS J1614-North\footnotemark[$*$] & 501042010 & \timeform{331.5717D} & \timeform{$-$0.5274D} & 2006/09/15 16:00 & 44.5 & - \\ \hline
\multicolumn{7}{@{}l@{}}{\hbox to 0pt{\parbox{85mm}{\footnotesize
      \par\noindent
      \footnotemark[$*$] \cite{Mat08}.
      \par\noindent
      }\hss}}
    \end{tabular}
  \end{center}
\end{table*}

\section{Analysis and Results}

\subsection{XIS Image}

We extracted XIS images from each sensor using the screened
data for the soft- and hard-energy bands.  For the FI
sensors, the soft- and hard-bands are defined as 0.4--3 keV
and 3--10 keV, respectively, while those for the BI sensor
are defined as 0.3--3 keV and 3--7 keV, respectively.  We
excluded the corners of the CCD chips illuminated by the
$^{55}$\rm{Fe} calibration sources.  The images of the
non-X-ray background (NXB) were generated using {\tt
  xisnxbgen}~\citep{Taw08} and subtracted from the HESS
J1614 images.  Then, the soft- and hard-band images were
divided by flat sky images simulated at 1.49 and 4.5 keV
using the XRT+XIS simulator {\tt xissim}~\citep{Ish07}
for vignetting corrections.  The images from the
two FI sensors were summed and binned by a factor of 8.

\begin{figure*}
  \begin{tabular}{cc}
    \begin{minipage}{0.5\hsize}
      \begin{center}
        \FigureFile(80mm,80mm){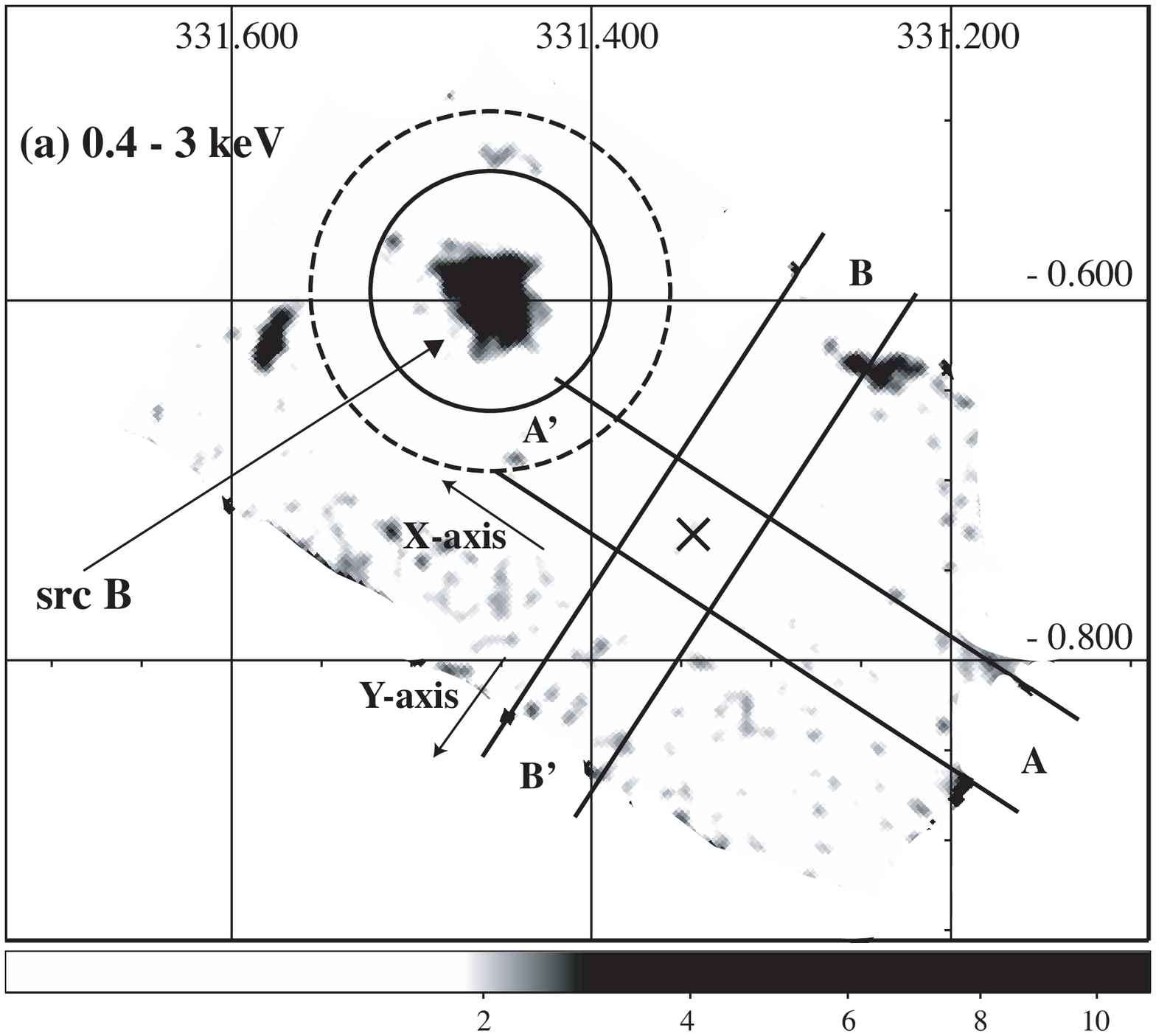}
      \end{center}
    \end{minipage}
    \begin{minipage}{0.5\hsize}
      \begin{center}
        \FigureFile(80mm,80mm){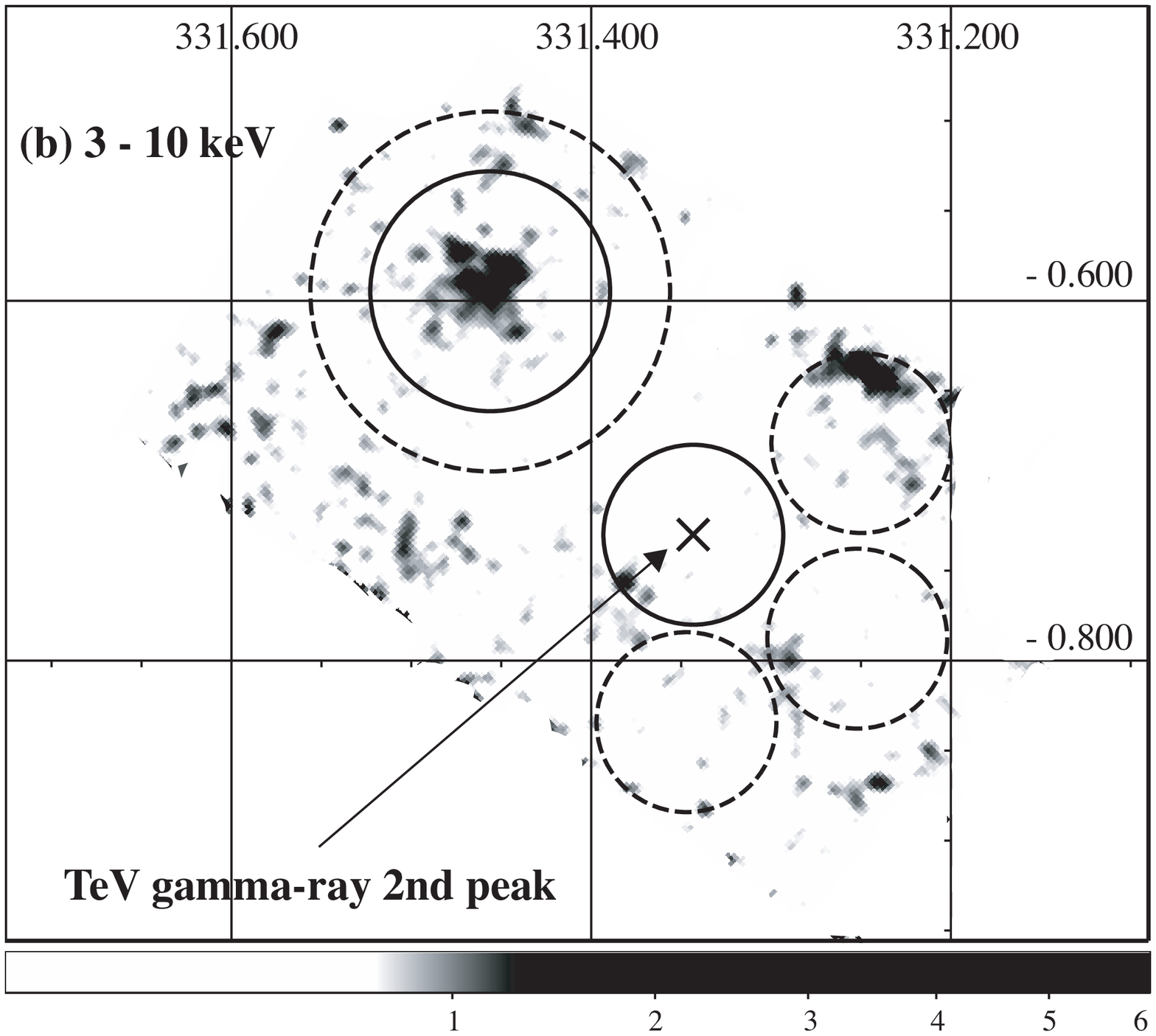}
      \end{center}
    \end{minipage}
  \end{tabular}
  \caption{Suzaku XIS FI (XIS 0+3) images of the HESS J1614 region (HESS J1614-South and HESS J1614-Center) in the Galactic coordinates: (a) 0.4--3 keV and (b) 3--10 keV bands. The images were smoothed using a Gaussian function with $\sigma$ = \timeform{0'.42}. A vignetting correction was applied after subtracting the NXB, as described in the text. The cross mark represents the position of the 2nd brightest peak of the TeV $\gamma$-ray emission. The solid lines in the left panel show the regions used for the photon count profiles shown in figure \ref{fig:3rd} (see section 3.1.1). The solid circle and the dashed circles in the right panel show the regions used for the determination of the upper limit to the X-ray emission (see section 3.2.1). The solid circle centered on src B is the integration region of source photons, and the dashed circle excluding the source region is that of background photons (see section 3.2.2). 
  }\label{fig:2nd}
\end{figure*}

The XIS FI images of the HESS J1614 region (HESS J1614-South
and HESS J1614-Center) shown in figure \ref{fig:2nd} were
smoothed using a Gaussian function with $\sigma$ =
\timeform{0'.42}.  The BI images were essentially the same,
except for the poorer statistics.  In the soft-band image, a
bright X-ray object with a peak position of ($l$, $b$) =
(\timeform{331.45D}, \timeform{$-$0.59D})
\footnote{($\alpha$, $\delta$)$_{\rm J2000.0}$
$=$(\timeform{16h14m04s}, $-$\timeform{51D52'27"}).} 
was found.  The position uncertainty, defined as a sigma of
a Gaussian function obtained by fitting the projection of
the object along the Galactic longitude, was
\timeform{0'.4}.  Thus, this object is coincident with
Suzaku J1614$-$5152 (src B).  Src B was also conspicuous in
the hard-band image.

\subsubsection{HESS J1614-South}

\begin{figure*}
  \begin{tabular}{cc}
    \begin{minipage}{0.5\hsize}
      \begin{center}
        \FigureFile(80mm,80mm){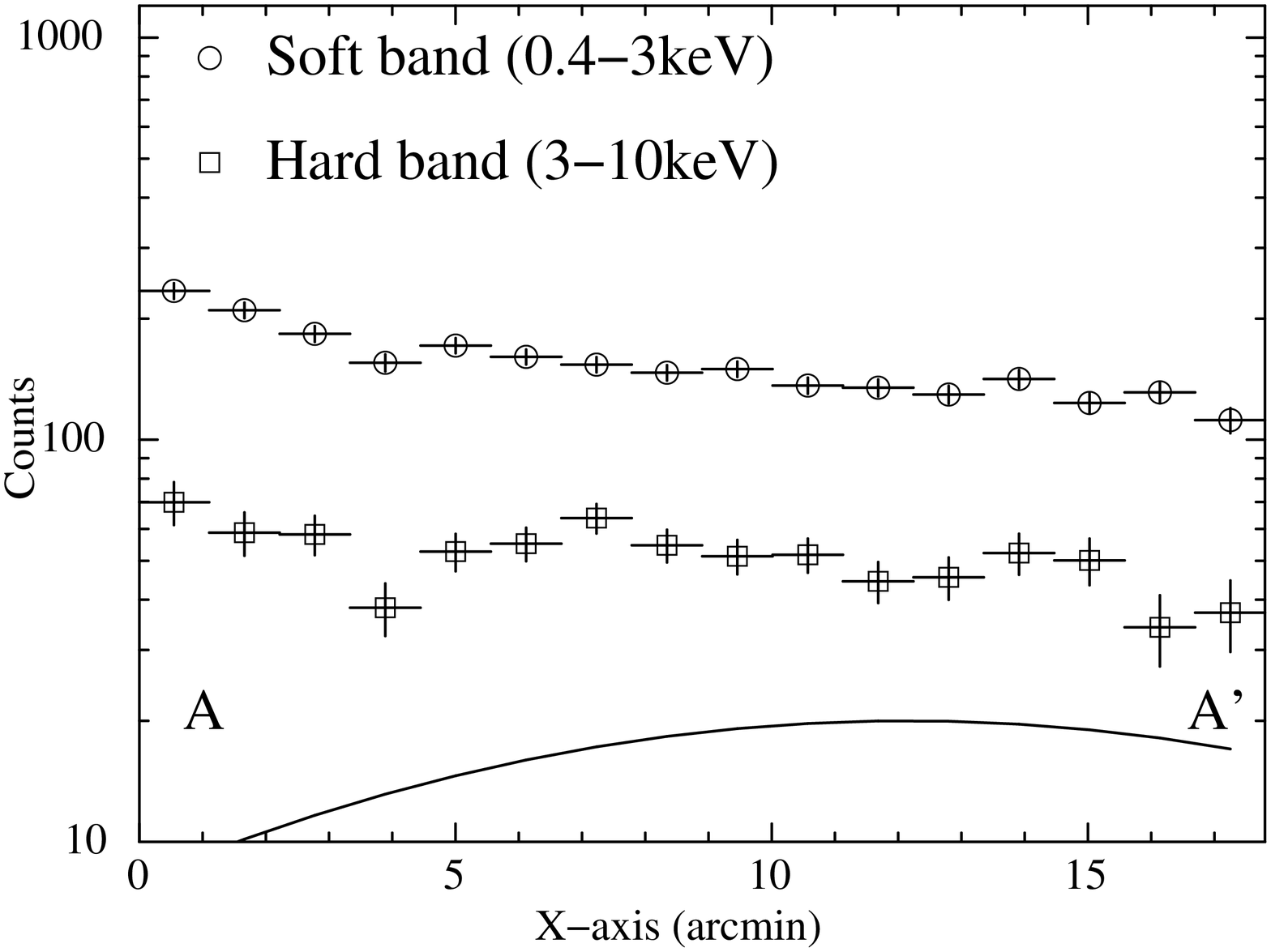}
      \end{center}
    \end{minipage}
    \begin{minipage}{0.5\hsize}
      \begin{center}
        \FigureFile(80mm,80mm){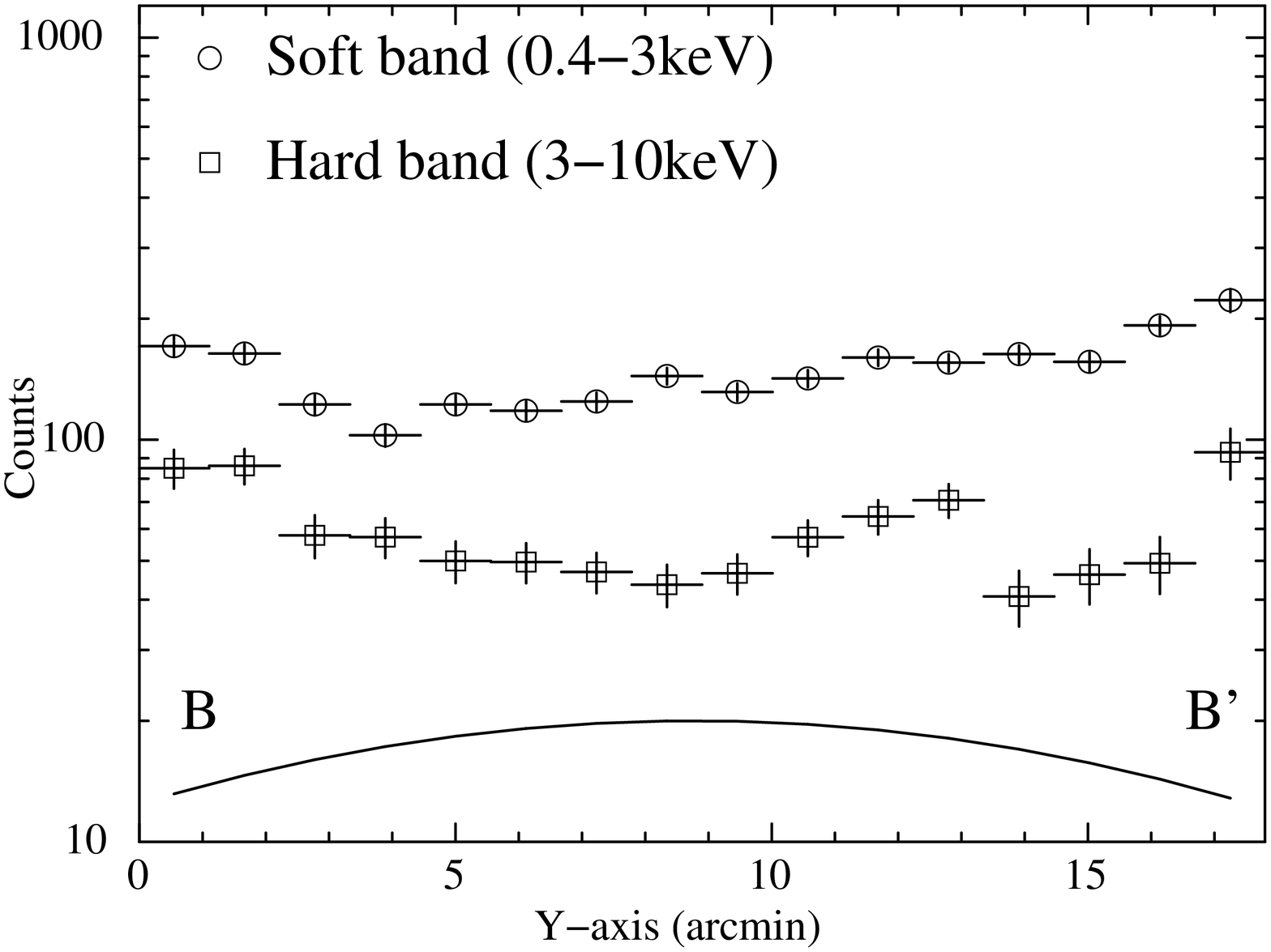}
      \end{center}
    \end{minipage}
  \end{tabular}
  \caption{Photon count profiles of the XIS images along the strips AA' (left) and BB' (right) shown in figure \ref{fig:2nd}(a). The solid curve shows the Gaussian function of $\sigma$ = \timeform{9'.0}, which expresses the TeV $\gamma$-ray profile of HESS J1614-South.
  }\label{fig:3rd}
\end{figure*}

There is no apparent X-ray structure suggesting an X-ray counterpart of the 2nd peak of the TeV $\gamma$-ray emission in either the soft- and hard-band images.
To examine this quantitatively, we made photon count profiles along the strips AA' and BB' with a width of \timeform{3'.7} in figure \ref{fig:2nd}(a); the profiles are shown in figure \ref{fig:3rd}.
We see no systematic trend, consistent with the TeV $\gamma$-ray profile of HESS J1614-South described by a Gaussian function with $\sigma$ = \timeform{9'.0}~\citep{Row08}, in either the soft and hard X-ray profiles.
These profiles strengthen the absence of the X-ray counterpart corresponding to the 2nd peak of the TeV $\gamma$-ray emission. 

\subsubsection{HESS J1614-Center}

\begin{figure}
  \begin{center}
    \FigureFile(80mm,80mm){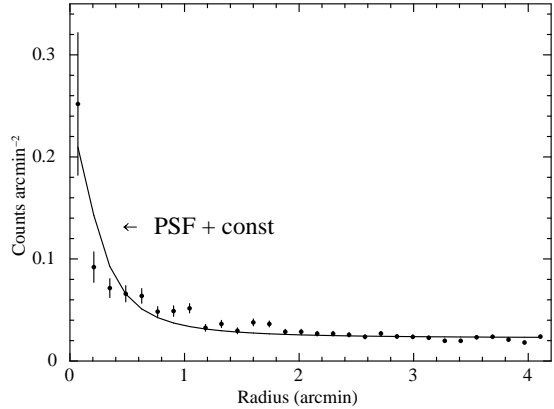}
  \end{center}
  \caption{Radial profile of src B extracted from the 3--10 keV band image of the XIS FI sensor (XIS 0+3). The solid line represents the XIS PSF profile with a constant component.}\label{fig:4th}
\end{figure}

We created the radial profile of src B and compared it with
a point-spread function (PSF).  The origin of the radial
profile of src B was the peak of the X-ray emission.  As for
the PSF, we obtained the radial profile using the SS Cyg
data observed on 2005 November 2 (OBSID=400006010), which
are the verification phase data for the imaging capability
of the XRT~\citep{Ser07}.  Since the energy
dependence of the PSF is negligible~\citep{Ser07},
the radial profile was extracted from the 0.4--10
keV band.  In this analysis, NXB subtraction and vignetting
correction were not applied to both the radial profile of src B
and the PSF.  Figure \ref{fig:4th} shows the radial profile
of src B in the 3--10 keV band.  The profile cannot be fitted with
the PSF plus a constant component model
($\chi^2$/d.o.f. = 99.20/28), and therefore src B must be
an extended source or unresolved multiple sources.

\subsection{XIS Spectrum}

\subsubsection{HESS J1614-South}

There was no apparent X-ray counterpart corresponding
to the 2nd peak of the TeV $\gamma$-ray emission, and we
estimated the upper limit of the X-ray flux of this region.
In order to determine the upper limit, we assumed a spectrum
similar to src A; an absorbed power-law with the photon
index $\Gamma = 2.0$ and the hydrogen-equivalent column
density $N_{\rm H} = 1.2 \times 10^{22}$ {\rm cm}$^{-2}$~\citep{Mat08}.
The cross sections of photoelectric absorption were obtained
from \citet{Mor83}.

We extracted the photons from the source region, the solid
circle with a radius of \timeform{3'} centered on the TeV
$\gamma$-ray 2nd peak in figure \ref{fig:2nd}(b), in the
2--10 keV band using the FI sensors (XISs 0 and 3).  The number
of the accumulated photons was
$1111 \pm 33$ counts.
Using three source-free regions surrounding the source
region, the dashed circles around the TeV $\gamma$-ray 2nd
peak in figure \ref{fig:2nd}(b), we estimated the photons of
the background emission.  The number of the accumulated
photons scaled the area to that of the source region was
$1194 \pm 35$ counts.  The exposure of the source was
the same as that of the background.  Thus, the number of the source minus
background photons was $-83 \pm 48$ counts, and the
3$\sigma$ confidence upper limit on the photons in the
source region was estimated to be 61 counts.  Therefore, the
3$\sigma$ confidence upper limit on the count rate in the
source region was $1.1 \times 10^{-3}$ counts s$^{-1}$.
Then, we calculated the flux upper limit with Redistribution
Matrix Files (RMFs) and Ancillary Response Files (ARFs).  We
obtained RMFs using the {\tt xisrmfgen}, and made ARFs for a
flat emission with the {\tt xissimarfgen}
software~\citep{Ish07}.  The 3$\sigma$ upper limit on the
surface brightness is $5.6 \times 10^{-15}$ {\rm erg
  cm$^{-2}$ s$^{-1}$ arcmin$^{-2}$}.  Therefore, the upper
limit on the flux from the circle of \timeform{3'} radius
(figure \ref{fig:2nd}(b)) is $1.6 \times 10^{-13}$ {\rm erg
  cm$^{-2}$ s$^{-1}$}.

\subsubsection{HESS J1614-Center}

A source region for src B is defined as the solid circle
centered on src B in figure~\ref{fig:2nd}.
We extracted light curves of src B in the soft- and hard-energy bands and
found no significant time variability from them.  Then, we
extracted the XIS spectra of src B and subtracted 
background spectra.  The source photons were extracted
from the source region, whereas those of the background
were extracted from the
dashed circle excluding the source region (figure \ref{fig:2nd}).  
We then combined
the background-subtracted spectra obtained from the two FI
sensors.
We obtained RMFs and ARFs for a point source
using the {\tt xisrmfgen} and {\tt xissimarfgen} software.

\begin{figure}
  \begin{center}
    \FigureFile(80mm,80mm){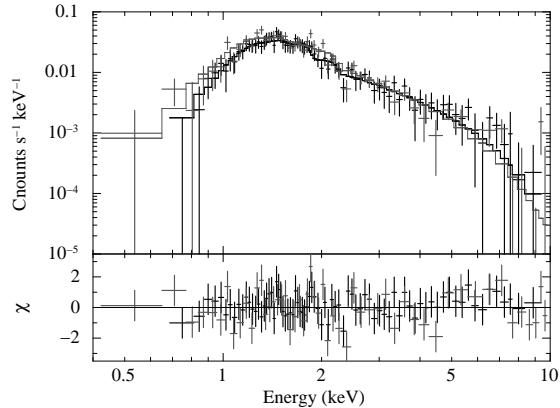}
  \end{center}
  \caption{XIS spectra of src B, shown with the best-fit power-law model. Black and gray lines represent the data and model for the XIS FI (XIS0+3) and XIS BI (XIS1), respectively.}\label{fig:5th}
\end{figure}

The spectra of src B are shown in figure \ref{fig:5th}.  We
fitted the spectra with an absorbed power-law model.  The
hydrogen-equivalent column density $N_{\rm H}$, the photon
index $\Gamma$, and the normalization were set to be free
parameters.  The best-fit parameters are $N_{\rm H} =
1.1^{+0.2}_{-0.1} \times 10^{22}$ {\rm cm}$^{-2}$ and
$\Gamma = 3.2^{+0.3}_{-0.2}$.  The best-fit $\chi^2$ value
is 103.47 for 128 degrees of freedom.  The observed flux in
the 2--10 keV band is $F$(2--10 keV) = $5.2 \times 10^{-13}$
{\rm erg cm$^{-2}$ s$^{-1}$}.

We also tried fitting an absorbed blackbody model.  The
following are the best-fit parameters: $N_{\rm H} =
0.29^{+0.10}_{-0.09} \times 10^{22}$ {\rm cm}$^{-2}$ and $kT
= 0.52^{+0.04}_{-0.03}$ keV.  However, the best-fit $\chi^2$
value of 169.24 for 128 degrees of freedom rejected the
validity of the blackbody model at a confidence level of
99\%.

\subsection{HXD Data Analysis}

We also analyzed the HXD-PIN data of HESS J1614-Center,
which includes src B.  The HXD-PIN spectrum of HESS
J1614-Center was extracted after the selection of good time
intervals.  As for the selection, we made a new GTI by
ANDing the GTI from the NXB file provided by the HXD team
with the GTI in the screened event file.  Dead-time
correction was also applied to the extracted source
spectrum.  Since the NXB from charged particles modeled by
the HXD team does not contain the cosmic X-ray background
(CXB) component~\citep{Bol87}, we simulated the CXB for HESS J1614-Center
following a recipe\footnote{See
  $\langle$http:\slash\slash{}heasarc.nasa.gov\slash{}docs\slash{}suzaku\slash{}analysis\slash{}pin\_cxb.html$\rangle$.}
provided by the HXD team.  We also used the response file
for Epoch 5 provided by the HXD team.  Note that the
accuracy of the HXD-PIN background model was estimated to be
less than 3\% in the 15--40 keV range~\citep{Fuk09}.

The observed HXD-PIN count of HESS J1614-Center in the
15--40 keV band was $11559 \pm 108$ counts.  The NXB and the
simulated CXB count in the 15--40 keV band were $10432 \pm
102$ counts and $642 \pm 25$ counts, respectively.  Thus,
the observed HXD-PIN minus NXB+CXB count was $485 \pm 150$
counts.  However, considering the accuracy of the HXD-PIN
background model of 3\%, the lower limit on the source count
of HESS J1614-Center became $172 \pm 151$ counts, and the
detection was at only 1.1$\sigma$.  There are not enough net
counts to study the HXD-PIN spectrum.  We also searched a
pulsation in the HXD-PIN data, but we did not detect any
pulsation in the 15--40 keV band.

\subsection{XMM-Newton Analysis}

The XMM-Newton observation of the HESS J1614 region was
carried out from 2007 February 13 17:54 (UT) to February 14
03:07 (UT) (OBSID=0406550101), and this observation covered
the HESS J1614-Center region with the EPIC instrument, which
consists of one pn-type CCD camera~\citep{Str01}
and two MOS CCD cameras~\citep{Tur01}.  In our Suzaku
analysis, the radial profile of src B indicates that src B
must be an extended source or unresolved multiple sources,
and the light curves of src B do not show the time
variability.  The EPIC instrument provides a high spatial
resolution, whose PSF (half energy width) is $\sim$
\timeform{15"}.  Moreover, the time resolutions of each EPIC
instrument are much better than the Suzaku/XIS (8~s),
73.4~ms for the pn and 2.6~s for the MOS in the full-frame
mode.  We therefore analyzed the XMM-Newton archival data.

The pn, MOS1 and MOS2 were operated in the standard
full-frame mode using the medium filter.  We used the
Standard Analysis System (SAS) software version 10.0.0 for
event selection.
We selected the photons with the PATTERN of 0--4 for the pn
and those of 0--12 for the MOS1 and MOS2, as valid X-ray
events.  We removed the time intervals in which the count
rate in the 10--12 keV band within the entire field of view
was higher than 0.25~counts~s$^{-1}$ for the pn and
0.20~counts~s$^{-1}$ for the MOS1 and MOS2, since such
events are considered to be a rapid increase in the
background induced by soft protons.  The resultant exposure
time was 1 ks for the pn and 10 ks for the MOS1 and MOS2.
Since the photon statistics were poor for the pn, we did not
use the pn data in this analysis.

\subsubsection{Image Analysis}

\begin{figure*}
  \begin{center}
    \FigureFile(120mm,80mm){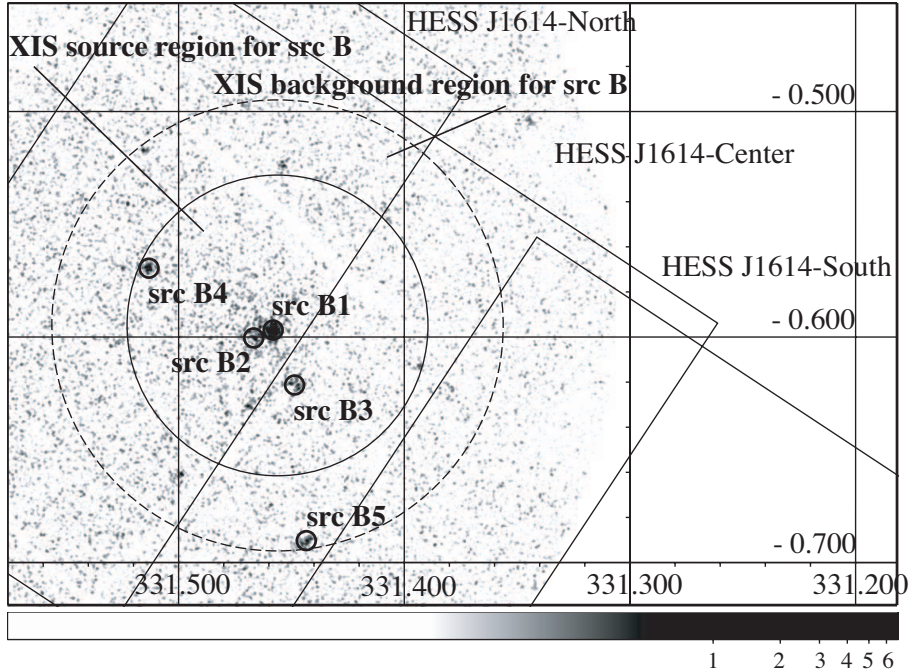}
  \end{center}
  \caption{MOS 1+2 zoomed image in the 0.4--10 keV band including the XIS source (solid circle) and background (dashed circle excluding the source region) regions for src B. The solid squares represent the Suzaku/XIS field of views of the HESS J1614-North/HESS J1614-Center/HESS J1614-South observations. We also show the extraction regions for the detected point sources, which are used to estimate the source count rates, with thick open circles. Lines of constant Galactic latitude and longitude are plotted and labeled in the interior of the figure.
  }\label{fig:6th}
\end{figure*}

Figure \ref{fig:6th} shows the combined MOS1 and MOS2 image
in the 0.4--10 keV band.  We found several sources within
the source and background regions of src B used to create
the XIS spectra.  We therefore searched for X-ray point
sources using the source detection task {\tt edetect\_chain}
in the SAS software package.  We set the lower threshold of
maximum likelihood method used in {\tt edetect\_chain} to be
10 (corresponding roughly to 4$\sigma$ detection).  The
detected point sources in the 0.4--10 keV band are shown
with thick open circles in figure \ref{fig:6th}.  Four point
sources were detected in the XIS source region for src B,
while only one point source was detected in the XIS
background region for src B.  We designated these objects as
XMMU J161406.0$-$515225 (src B1), XMMU J161409.3$-$515213
(src B2), XMMU J161409.8$-$515352 (src B3), XMMU
J161414.1$-$514857 (src B4), and XMMU J161426.8$-$515705
(src B5).  For each point source, we accumulated photons
from a circular region with a radius of \timeform{15"}, and
then estimated their count rates.  Table \ref{tab:2nd}
summarizes the count rates of these detected point sources.
We also gave the count rates of the XIS source and
background regions for src B excluding the point sources in
order to indicate the count rate of the background photons
in this field.  The count rate of src B1 is higher than the
other sources by a factor of $\sim 5$.  Moreover, the peak
position of src B1, ($l$, $b$) = (\timeform{331.45D},
\timeform{$-$0.59D})
\footnote{($\alpha$, $\delta$)$_{\rm J2000.0}$
$=$(\timeform{16h14m06s}, $-$\timeform{51D52'25"}).} 
with the position uncertainty
of \timeform{2".8}, is consistent with that of Suzaku src B.
These indicate that Suzaku src B is multiple sources and the
main object is src B1.

\begin{table*}[htbp]
  \begin{center}
    \caption{Count rates of XMM-Newton data.}\label{tab:2nd}
    \begin{tabular}{lcccccc}\hline
  & & & & \multicolumn{3}{c}{Count rate\footnotemark[$*$]} \\
  & & & & \multicolumn{3}{c}{[10$^{-3}$~counts~s$^{-1}$]} \\ \cline{5-7}
Region & R.A. (J2000.0) & Decl. (J2000.0) & & 0.4 -- 3 keV & 3 -- 10 keV & Total \\ \hline
src B1\footnotemark[$\dagger$] & 16 14 06 & $-$51 52 25 & & 22.7$\pm$1.0 & 2.03$\pm$0.31 & 24.7$\pm$1.1 \\
src B2\footnotemark[$\dagger$] & 16 14 09 & $-$51 52 13 & & 4.70$\pm$0.48 & 0.629$\pm$0.175 & 5.33$\pm$0.51 \\
src B3\footnotemark[$\dagger$] & 16 14 09 & $-$51 53 52 & & 3.59$\pm$0.42 & 0.437$\pm$0.146 & 4.03$\pm$0.44 \\
src B4\footnotemark[$\dagger$] & 16 14 14 & $-$51 48 57 & & 4.66$\pm$0.48 & 0.776$\pm$0.194 & 5.43$\pm$0.51 \\ 
src B5\footnotemark[$\dagger$] & 16 14 26 & $-$51 57 05 & & 3.83$\pm$0.43 & 0.242$\pm$0.108 & 4.08$\pm$0.44 \\ \hline
\multicolumn{3}{l}{XIS source region $-$ (src B1 $+$ src B2 $+$ src B3 $+$ src B4)\footnotemark[$\ddagger$]} & & 0.670$\pm$0.011 & 0.405$\pm$0.009 & 1.07$\pm$0.01 \\ \hline
\multicolumn{3}{l}{XIS background region $-$ src B5\footnotemark[$\ddagger$]} & & 0.563$\pm$0.009 & 0.377$\pm$0.008 & 0.940$\pm$0.012 \\ \hline
\multicolumn{7}{@{}l@{}}{\hbox to 0pt{\parbox{180mm}{\footnotesize 
      \par\noindent
\\
\footnotemark[$*$] All errors are at the 1$\sigma$ confidence level.
\par\noindent
\footnotemark[$\dagger$] Extraction regions are the circle with a radius of \timeform{15"}.
\par\noindent
\footnotemark[$\ddagger$] Extraction regions are scaled to the circle with a radius of \timeform{15"}.
    }\hss}}
    \end{tabular}
  \end{center}
\end{table*}

\begin{figure}
  \begin{center}
    \FigureFile(80mm,80mm){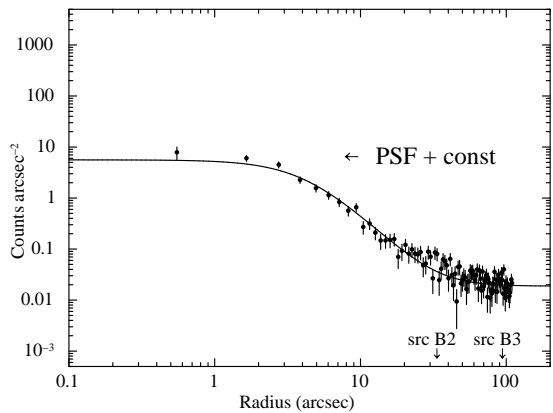}
  \end{center}
  \caption{Radial profile of src B1 extracted from the MOS detectors in the 1.2--2.4 keV band. The solid line represents the PSF profile with a constant component. The downward arrows represent the positions of src B2 and src B3 (see figure \ref{fig:6th})}\label{fig:7th}
\end{figure}

Figure \ref{fig:7th} shows the radial profile of src B1
extracted from the MOS detectors in the 1.2--2.4 keV band;
The origin of the radial profile was the peak of the X-ray
emission, and NXB subtraction and vignetting correction were
not applied to it.
Then, we compared it with a point-spread function (PSF) in order to investigate whether or not src B1 is a point source. 
Since the PSF of the XMM-Newton telescopes depends on the
energy, we used the PSF at an energy of 1.8 keV.  We
parameterized the PSF at 1.8~keV with a King-type function:
${\rm PSF}(r)=A(1+(r/r_{\rm c})^{2})^{\alpha}$, where $A$ is
a normalization, $r_{\rm c}$ = \timeform{4".36} is a core
radius and $\alpha = -1.41$ is a slope (see the
XMM-SOC-CAL-TN-0022 and XMM-SOC-CAL-TN-0029 documents online
at http://xmm.vilspa.esa.es).  We fitted the radial profile
in the range of \timeform{0".1} $< r <$ \timeform{200"} with
the above PSF model plus constant which represents the
background.  In this fit, the normalization and the constant
were free.  We overlaid the best-fit model on the radial
profile in figure \ref{fig:7th}.  The best-fit $\chi^2$
value is 109.6 for 98 degrees of freedom.  This result
suggests that src B1 is a point source.

\subsubsection{Spectral Analysis}

\begin{figure}
  \begin{center}
    \FigureFile(80mm,80mm){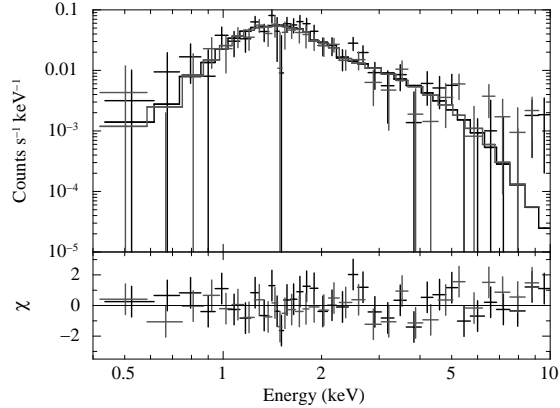}
  \end{center}
  \caption{MOS1 (black) and MOS2 (grey) spectra of src B. The data and their best-fit power-law model are represented by cross marks and solid lines, respectively.
  }\label{fig:8th}
\end{figure}

In order to check for consistency with the Suzaku results,
we extracted the MOS spectra of the XIS source region for
src B and subtracted the background spectra of the XIS
background region for src B (figure \ref{fig:6th}).  We
fitted the spectra with an absorbed power-law model.  The
fit is acceptable with the best-fit $\chi^2$ value of 47.98
for 65 degrees of freedom.  The best-fit parameters are
$N_{\rm H} = 1.1^{+0.4}_{-0.3} \times 10^{22}$ {\rm
  cm}$^{-2}$ and $\Gamma = 3.0^{+0.6}_{-0.5}$.  The observed
flux in the 2--10 keV band is $F$(2--10 keV) = $6.9 \times
10^{-13}$ {\rm erg cm$^{-2}$ s$^{-1}$}.  Figure
\ref{fig:8th} shows the background-subtracted spectra
together with the best-fit model and the fit residuals.  The
best-fit results of Suzaku src B and XMM-Newton src B are
summarized in table \ref{tab:3rd}.  The best-fit parameters
of src B are consistent with Suzaku and XMM-Newton.
The X-ray flux of XMM-Newton is $\sim$30\% larger than
that of Suzaku, and this may indicate time variability.

\begin{table*}[htbp]
  \begin{center}
    \caption{Best-fit results of the X-ray spectra.\footnotemark[$*$]}\label{tab:3rd}
    \begin{tabular}{lcccc}\hline
 & Suzaku src B & XMM-Newton src B & \multicolumn{2}{c}{XMM-Newton src B1} \\ \hline
Model\footnotemark[$\dagger$]  & PL & PL & PL & BB \\ 
$N_{{\rm H}}$ ($10^{22}{\rm cm}^{-2}$) & $1.1^{+0.2}_{-0.1}$ & $1.1^{+0.4}_{-0.3}$ & $2.4^{+0.4}_{-0.4}$ & $1.1^{+0.3}_{-0.2}$ \\ 
$\Gamma$/(kT) (keV) & $3.2^{+0.3}_{-0.2}$ & $3.0^{+0.6}_{-0.5}$ & $5.2^{+0.6}_{-0.5}$ & $0.38^{+0.04}_{-0.04}$ \\ 
$F^{{\rm obs}}_{2-10{\rm keV}}$\footnotemark[$\ddagger$] & $5.2$ & $6.9$ & $1.7$ & $1.5$ \\ 
$F^{{\rm abscor}}_{2-10{\rm keV}}$\footnotemark[$\S$] & $6.1$ & $8.0$ & $2.8$ & $2.0$ \\ 
$\chi^{2}$/d.o.f. & $103.47/128$ & $47.98/65$ & $21.98/25$ & $24.91/25$ \\ \hline
\multicolumn{5}{@{}l@{}}{\hbox to 0pt{\parbox{140mm}{\footnotesize
      \par\noindent
\\
\footnotemark[$*$] Errors are at the 90\% confidence level.
\par\noindent
\footnotemark[$\dagger$] Model used for the spectral fitting: "PL" is a power-law model, and "BB" is a blackbody model.
\par\noindent
\footnotemark[$\ddagger$] Observed flux in the 2--10 keV band in units of $10^{-13}$ {\rm erg cm$^{-2}$ s$^{-1}$}.
\par\noindent
\footnotemark[$\S$] Absorption corrected flux in the 2--10 keV band in units of $10^{-13}$ {\rm erg cm$^{-2}$ s$^{-1}$}.
\par\noindent
    }\hss}}
    \end{tabular}
  \end{center}
\end{table*}

\begin{figure}
  \begin{center}
    \FigureFile(80mm,80mm){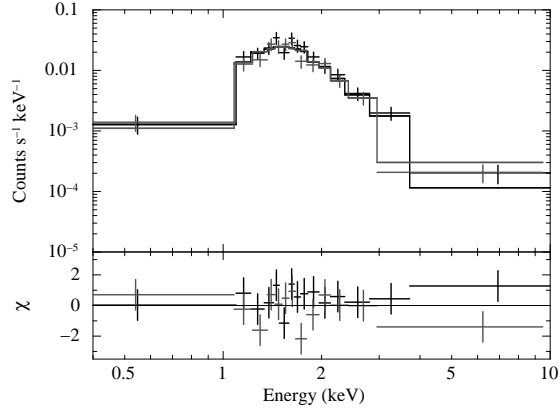}
  \end{center}
  \caption{MOS1 (black) and MOS2 (grey) spectra of src B1. The data and their best-fit power-law model are represented by cross marks and solid lines, respectively.
  }\label{fig:9th}
\end{figure}

Then, we extracted the MOS spectra of src B1.  The circle
with a radius of \timeform{20"} centered on src B1 was the
extraction region of the source photons, whereas the XIS
background region for src B excluding the circle with a
radius of \timeform{20"} centered on src B5 was that of the
background photons (figure \ref{fig:6th}).  The spectra of
src B1 are shown in figure \ref{fig:9th}.  We fitted the
spectra with an absorbed power-law model.  The best-fit
parameters are $N_{\rm H} = 2.4^{+0.4}_{-0.4} \times
10^{22}$ {\rm cm}$^{-2}$ and $\Gamma = 5.2^{+0.6}_{-0.5}$.
We also tried to fit the spectra by an absorbed blackbody
model.  The best-fit parameters are as follows: $N_{\rm H} =
1.1^{+0.3}_{-0.2} \times 10^{22}$ {\rm cm}$^{-2}$ and $kT =
0.38^{+0.04}_{-0.04}$ keV.  The best-fit parameters of the
power-law and blackbody models are listed in table
\ref{tab:3rd}.  Moreover, we tried fitting a thermal plasma
model (the MEKAL model:~\cite{Mew85}).  However, the
thermal model yielded an abundance of zero ($< 0.044$)
solar, which is an extremely low abundance, and therefore is
not realistic.  Since the statistics were poor, it was impossible
to study the spectra of the other sources.

\subsubsection{Timing Analysis}

\begin{figure}
  \begin{center}
    \FigureFile(80mm,80mm){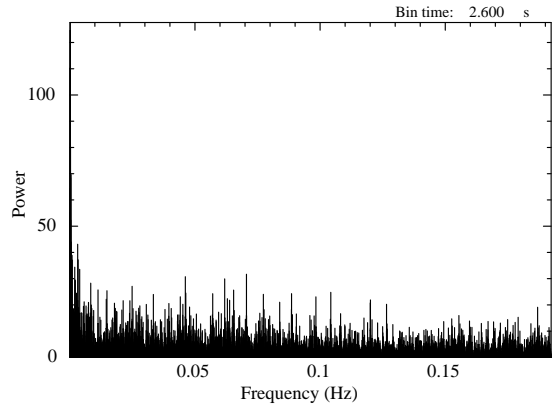}
  \end{center}
  \caption{Power spectrum of src B1 in the 1--5 keV band with the time bin size of 2.6 s for the MOS data.}\label{fig:10th}
\end{figure}

We carried out a timing analysis of src B1.  Since there was
nearly no source flux below $\sim$ 1 keV and above $\sim$ 5
keV, we searched a pulsation in the 1--5 keV band for the
MOS data.  After barycentric correction to the event file,
we created the light curve of src B1 with the minimum time
resolution (2.6~s).  The power spectrum on the basis of the
combined MOS1 and MOS2 light curve in the 1--5 keV band
is shown in figure \ref{fig:10th}.  We did not detect a
pulsation of src B1 for the MOS data.

\section{Discussion}

\subsection{The TeV-to-X-ray flux ratio of HESS J1614}

HESS J1614 has two regions with intense $\gamma$-ray
emission.  The $\gamma$-ray spectrum yields $\Gamma$ = 2.46
and the flux in the 1--10 TeV band from a circular region
with a radius of \timeform{0.4D} is estimated to be $1.8
\times 10^{-11}$ {\rm erg cm$^{-2}$ s$^{-1}$}~\citep{Aha06}.  \citet{Mat08}
found the hard extended
emission, Suzaku src A, at the position of the 1st brightest
peak of the TeV $\gamma$-ray emission, as the best candidate
for the X-ray counterpart of HESS J1614; the best-fit
parameters are $N_{\rm H} = 1.2^{+0.5}_{-0.4} \times
10^{22}$ {\rm cm}$^{-2}$ and $\Gamma = 1.7^{+0.3}_{-0.3}$,
and the flux of Suzaku src A in the 2--10 keV band is
$F$(2--10 keV) = $5.3 \times 10^{-13}$ {\rm erg cm$^{-2}$
  s$^{-1}$}.  Additionally, we reveal that there is no
evidence for the X-ray emission at the position of the 2nd
brightest peak of the TeV $\gamma$-ray emission; the XIS
sets the stringent upper limit of $1.6 \times 10^{-13}$ {\rm
  erg cm$^{-2}$ s$^{-1}$} in the 2--10 keV band.  As a
result, the X-ray counterpart to the TeV $\gamma$-ray
emission is just Suzaku src A.  Therefore, it is ensured
that the flux ratio $F$(1--10 TeV)/$F$(2--10 keV) of HESS
J1614 is $\sim 34$, which is still one of the largest values
observed among extended VHE objects (see \cite{Mat07} and references therein).

\subsection{The nature of HESS J1614}

XMM-Newton src B1 is located at the middle of HESS J1614.
Within the error circle of XMM-Newton src B1, there is an
infrared source, 2MASS J16140610$-$5152264, at ($l$, $b$) =
(\timeform{331.4579D}, \timeform{$-$0.5973D}).  The X-ray
spectrum of XMM-Newton src B1 is described using a power-law
model with $\Gamma = 5.2^{+0.6}_{-0.5}$ or a blackbody model
with $kT = 0.38^{+0.04}_{-0.04}$ keV.  The best-fit column
densities of both models are different by a factor of $\sim
2$.  The best-fit column density of the power-law model is
almost the same as the total Galactic HI column density
towards the HESS J1614 region ($\sim 2.2 \times 10^{22}$
{\rm cm}$^{-2}$:~\cite{Dic90}), while that of the
blackbody model is approximately equal to half of it.  On
the other hand, the best-fit column density of the power-law
model is about twice larger than that of Suzaku src A, while
that of the blackbody model is almost the same as that of
Suzaku src A.  If we assume the blackbody emission,
XMM-Newton src B1 may be at about the same distance to
Suzaku src A and may also be physically related to HESS
J1614.

An object which shows a soft X-ray spectrum like XMM-Newton
src B1 includes an anomalous X-ray pulsar (AXP
:~\cite{Mer08}).  The spectra of AXPs are described by steep power
laws ($\Gamma >$ 3) and/or blackbody models with low
temperature ($kT \leq$ 0.5 keV). AXPs often show time
variability with a period of 2--12~s. However, no such
variability was found from XMM-Newton src B1.

If XMM-Newton src B1 is related to Suzaku src A and HESS
J1614, the nature of HESS J1614 is similar to that of 
CTB37B~\citep{Nak09}.  CTB37B is an SNR which has the
diffuse non-thermal power-law component ($\Gamma \sim 1.5$)
at the position of the TeV $\gamma$-ray emission.  Moreover,
the point source which shows the soft spectrum considered to
be an AXP was found at the offset position of the TeV
$\gamma$-ray emission~\citep{Nak09,Sat10}.  Thus, if
XMM-Newton src B1 is an AXP, HESS~J1614
may be an SNR associated with an AXP.  In that case, it is
considered that XMM-Newton src B1 is an AXP produced by a
supernova explosion, and a shocked region in the SNR can be
seen as Suzaku src A.  The approximate radius of HESS J1614
($r \sim$ \timeform{10'} = 30~pc @ 10~kpc) suggests that
HESS J1614 may be an SNR with an age of $\sim$ $10^5$
yrs~\citep{Pad01}.  \citet{Yam06}
suggested that the ratio of TeV $\gamma$-ray to X-ray energy
flux of old SNRs with an age of $\sim 10^5$ yr could be very
large, in some instances more than $\sim 100$.  The flux
ratio of HESS J1614 is in agreement with this scenario.  Here,
we assume the distance to HESS J1614 is 10 kpc, since
the best-fit column density of the blackbody model which is
almost the same as that of Suzaku src A, is approximately
equal to half of the total Galactic HI column density
towards the HESS J1614 region.  Assuming a distance of 10 kpc,
the luminosities of XMM-Newton src B1 in the 2--10 keV band are
$3 \times 10^{33}$ {\rm erg s$^{-1}$} for the power-law model and
$2 \times 10^{33}$ {\rm erg s$^{-1}$} for the blackbody model.  These
luminosities are consistent with that of AXPs (e.g., The McGill SGR/AXP Online Catalog.\footnote{See $\langle$http:\slash\slash{}www.physics.mcgill.ca\slash{}$\sim$pulsar\slash{}magnetar\slash{}main.html$\rangle$.}). Meanwhile,
assuming a distance of 10 kpc, the luminosity of Suzaku src A
in the 2--10 keV band is $6 \times 10^{33}$
{\rm erg s$^{-1}$}~\citep{Mat08}.  This luminosity is smaller than
the synchrotron X-ray luminosity of young SNRs (e.g., \cite{Dye04}).

On the other hand, \citet{Row08} suggested the
relation between HESS J1614 and Pismis 22.  Pismis 22 is an
open cluster characterized by the following cluster
fundamental parameters: E(B$-$V) color excess =
2.00$\pm$0.10 mag, distance = 1.0$\pm$0.4 kpc and Age =
40$\pm$15 Myr~\citep{Pia00}.  \citet{Row08}
pointed out that Pismis 22 may easily account for the TeV
luminosity if the cluster has a stellar wind luminosity of
ten B-type stars and 20\% of the stellar wind kinetic energy
is converted to the $\gamma$-ray.  The size of Pismis 22 is
\timeform{4'.0}, it is conceivable that XMM-Newton src B1 is
within this cluster.  However, according to the best-fit
column densities of both models, the distance to XMM-Newton
src B1 of 1.0 kpc is too close.  Moreover, the spectrum of
XMM-Newton src B1 cannot be described by the optically thin
thermal plasma model seen from some O and B-type
stars~\citep{Col07}.  There are other TeV $\gamma$-ray
sources, such as Cyg-OB2 (TeV J2032+4130), suspected the
relation with the open cluster~\citep{Aha02}.  However,
in the case of TeV J2032+4130, the size
of the extended X-ray emission is similar to that of the TeV
$\gamma$-ray emission~\citep{Hor07}. Thus there are
some differences with HESS J1614.

\section{Summary}

We observed the south and center regions of HESS J1614 with
Suzaku. There was no positive detection at the 2nd peak
position, and we set an upper limit of $1.6 \times 10^{-13}$
{\rm erg cm$^{-2}$ s$^{-1}$} to the 2--10 keV band flux. The
high value of $f_{\rm TeV}/f_{\rm X} \sim 34$ may suggest
that HESS J1614 is a proton accelerator. We also detected
the soft X-ray source, Suzaku J1614$-$5152 (Suzaku src B),
at the middle of HESS J1614. Using the XMM-Newton archival
data, we revealed that Suzaku src B consists of multiple
point sources. The brightest point source, XMMU
J161406.0$-$515225 (XMM-Newton src B1), shows a soft X-ray
spectrum. XMM-Newton src B1 might be an AXP and may be
physically related to Suzaku J1614$-$5141 (Suzaku src A),
which was found at the 1st peak position in the previous
observation.

\bigskip

The authors are grateful to Professors W.~Hoffman and S.~Funk for kindly providing the HESS image. We thank H.~Kunieda, K.~Ishibashi, A.~Furuzawa, and H.~Mori for their useful comments. We also thank all Suzaku members. MS is supported by Grant-in-Aid for Japan Society for the Promotion of Science (JSPS) Fellows, 23·5737. HM is supported by Grant-in-Aid for Scientific Research (B), 22340046. This work was partially supported by the Grant-in-Aid for Nagoya University Global COE Program, "Quest for Fundamental Principles in the Universe: from Particles to the Solar System and the Cosmos", from the Ministry of Education, Culture, Sports, Science and Technology of Japan.

\end{document}